\documentclass[aps,preprint,floats,showpacs,prb]{revtex4}

\usepackage{amsmath,amssymb,graphicx,epsf}

\begin{document}

\title{Lifshitz transition in two-dimensional spin-density wave models} 

\author{Jie Lin}
\affiliation{Materials Science Division, Argonne National Laboratory, Argonne, IL 60439}

\begin{abstract}
We argue that both \textit{pocket-disappearing} and \textit{neck-disrupting} types of Lifshitz transitions can be realized in two-dimensional spin-density wave models for underdoped cuprates, and study both types of transitions with impurity scattering treated in the self-consistent Born approximation. We first solve for the electron self-energy from the self-consistent equation, and then study the low-temperature electrical conductivity and thermopower. Close to the Lifshitz transition, the thermopower is strongly enhanced. For the pocket-disappearing type, it has a sharp peak, while for the neck-disrupting type, it changes sign at the transition, with its absolute value peaked on both sides of the transition. We discuss possible applications to underdoped cuprates. 
\end{abstract}
\date{\today}
\pacs{72.15.Jf, 75.30.Fv, 74.72.Kf}

\maketitle

\section{Introduction}

One of the cornerstones of metallic physics is the concept of the Fermi surface,\cite{Lifshitz60Usp,Abrikosov88} the surface in momentum space separating filled states from empty ones at zero temperature. Applying external perturbations, such as hydrostatic pressure, anisotropic strain, or a change in the chemical composition, the Fermi surface can change its shape or more drastically its topology. This change of the Fermi surface topology is usually referred to as a Lifshitz transition,\cite{Lifshitz60} and has a profound impact on material properties.\cite{Kaganov79,Varlamov89,Blanter94,Bruno94} Recently, the Lifshitz transition has invoked revived interest in the context of cuprates,\cite{LeBoeuf10} iron-arsenic superconductors,\cite{Liu10} cold-atom systems,\cite{Carr10} bi-layer graphene,\cite{Lemonik10} and heavy-fermion metals and organic conductors.\cite{Yamaji06,Imada10} 

\begin{figure}
\centering
\includegraphics[width=.8\textwidth]{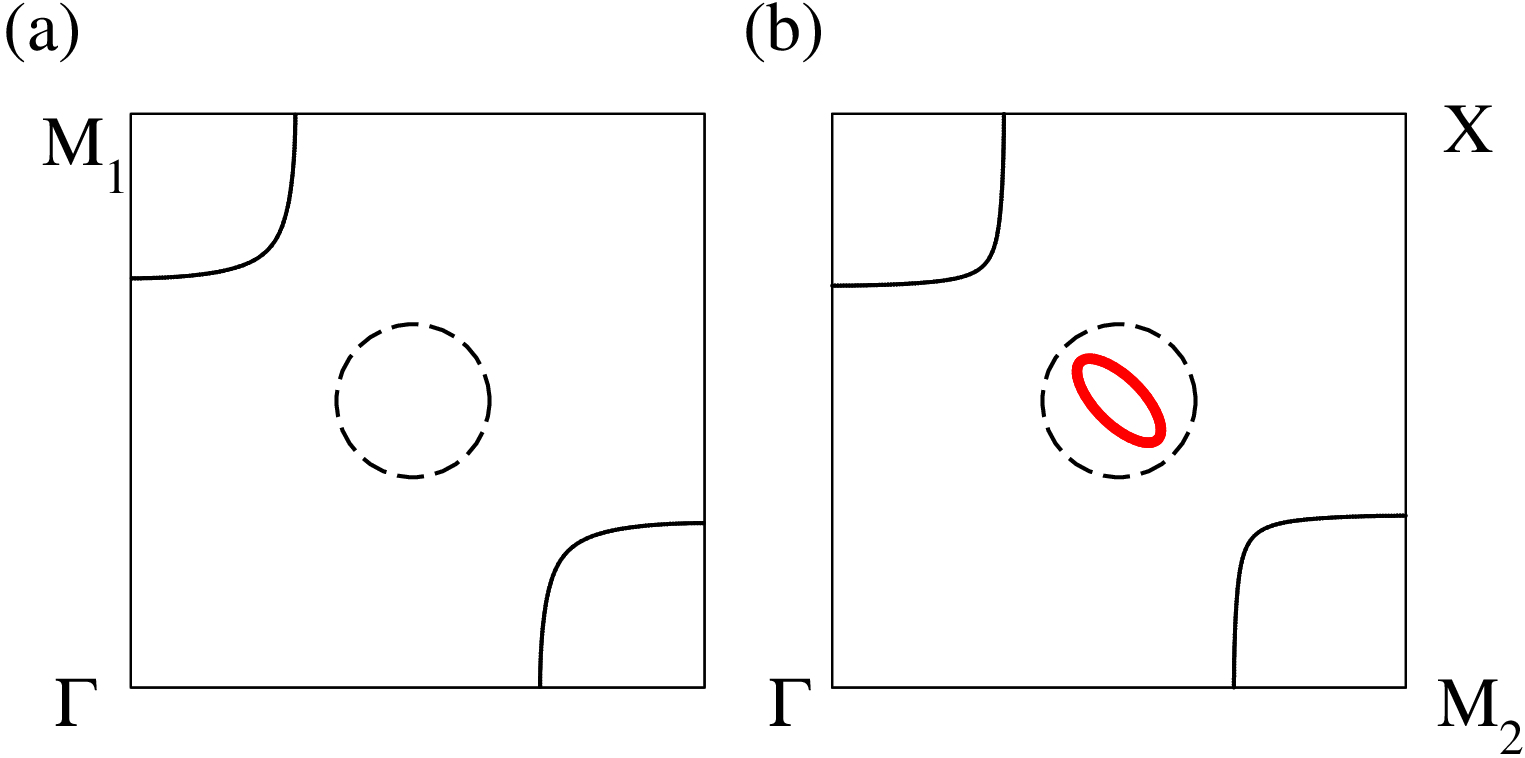}
\caption{\small{The Fermi surfaces for two different values of the spin potential in the commensurate spin-density wave model for electron-doped cuprates. This shows an example of the pocket-disappearing Lifshitz transition. Compared to (a), there is an extra small Fermi pocket represented by the thick solid line in (b). The momentum region where the new pocket forms is bounded by the dashed circle, and is defined as the $L$-region in the main text. In all Fermi surface plots in this paper, $\Gamma=(0,0)$, $X=(\pi,\pi)$, $M_1=(0,\pi)$, and $M_2=(\pi,0)$, where the planar lattice constant $a=1$.}}
\label{fig:pd}
\end{figure}

\begin{figure}
\centering
\includegraphics[width=.8\textwidth]{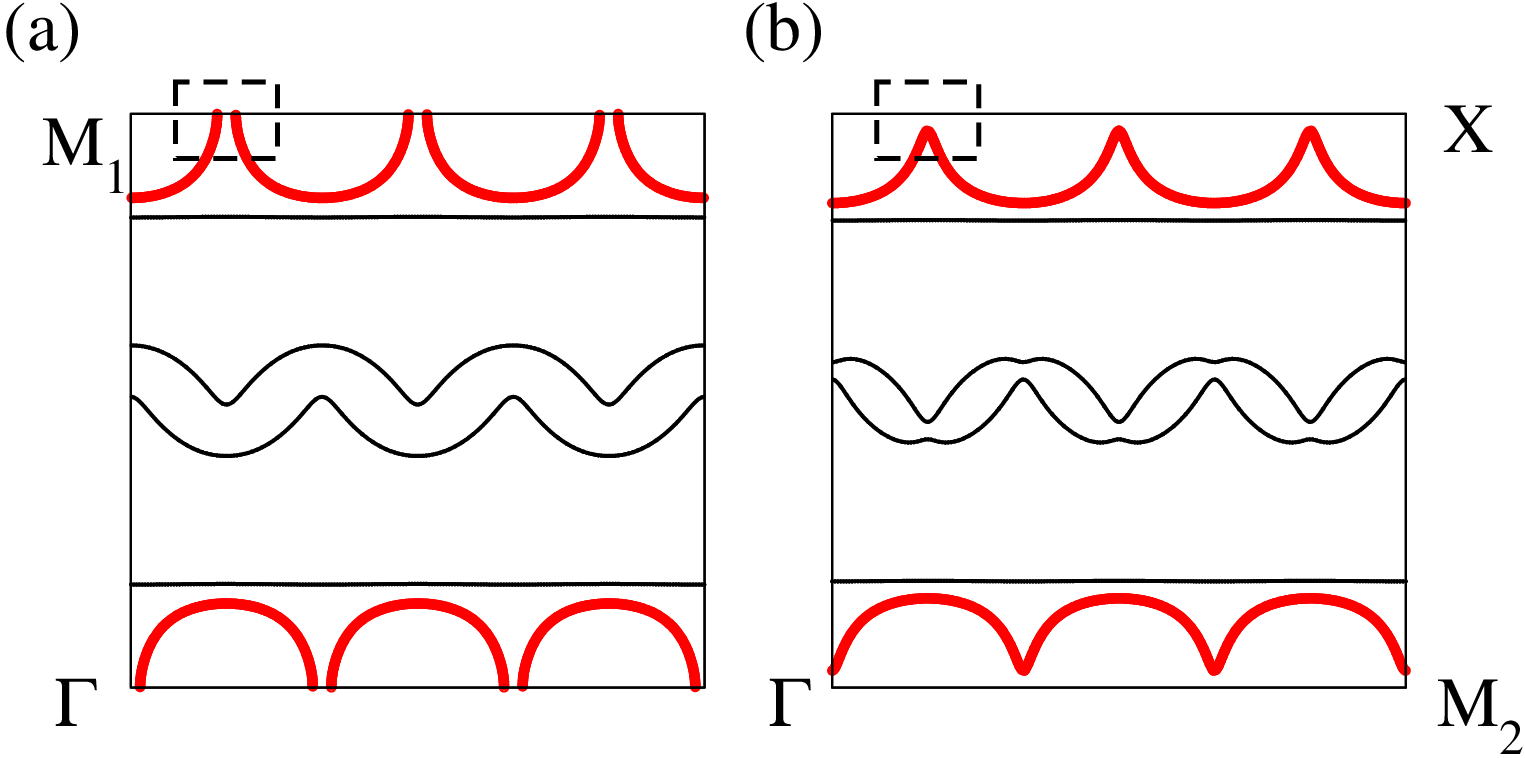}
\caption{\small{The Fermi surfaces for two different values of the stripe potential in the stripe model with ordering wave vector $(5\pi/6,\pi)$ for hole-doped cuprates. This shows an example of the neck-disrupting Lifshitz transition. From (a) to (b), the Fermi surface represented by the thick solid line changes its topology. The momentum region where this happens is bounded by the dashed box, and is defined as the $L$-region in the main text.}}
\label{fig:nd}
\end{figure}

It is known that a density wave instability can change the Fermi surface topology. In the density wave ordered state, the Fermi surface evolves with the strength of the coherent density wave potential.  Well-known examples are the underdoped cuprates. In the case of the electron-doped cuprates with commensurate $(\pi,\pi)$ spin-density wave order, it has been shown that the hole pockets around $(\pi/2,\pi/2)$ disappear as one tunes the magnitude of the density wave potential or the electron doping,\cite{Lin05} as shown in Figure \ref{fig:pd}. (We take the in-plane lattice constant $a=1$.) This is an example of the \textit{pocket-disappearing} Lifshitz transition.\cite{Lifshitz60} Another example of this type is found in the site-centered anti-phase stripe model, as shown in Refs [\onlinecite{Lin08,Lin09}], where both the electron pocket and the hole pockets can disappear independently by tuning the strength of the spin and charge potentials. This pocket-disappearing Lifshitz transition has important consequences for the Hall effect.\cite{Lin05,Lin08,Lin09} Recently, it was realized that the \textit{neck-disrupting} Lifshitz transition is also possible in a stripe model for the underdoped cuprates,\cite{Norman10} as shown in Figure \ref{fig:nd}, which provides a simple explanation for the disappearance of quantum oscillations and the associated diverging cyclotron mass in YBa$_2$Cu$_3$O$_{6+x}$.\cite{Sebastian10-1} 

In the single-particle picture, the energy dispersion of band electrons $\varepsilon_\mathbf{p}$ has extrema in the first Brillouin zone (local maxima, local minima, and/or saddle points) at energies denoted generically as $\epsilon_c$. As the chemical potential $\mu$ is varied through $\epsilon_c$, the Fermi surface changes its topology and a Lifshitz transition takes place; for a local maximum or minimum, a pocket appears or disappears, corresponding to the pocket-disappearing type transition, and for a saddle point, the Fermi surface changes its connection, corresponding to a neck-disrupting type transition. 

The cuprate band structure has saddle points located at $\mathrm{M}_{1,2}$ in Figure \ref{fig:pd}.\cite{Andersen95} Upon hole-doping, when $\mu$ reaches the energy of the saddle points, a neck-disrupting Lifshitz transition occurs, characterized by a logarithmically divergent density of states. There have been extensive studies on the van Hove singularity in the cuprate band dispersion, and its effects on material properties, as reviewed in Ref [\onlinecite{Markiewicz97}]. For the band structure obtained in Ref [\onlinecite{Andersen95}], this Lifshitz transition occurs at a relatively large hole doping. On the other hand, in the spin-density wave model, the neck-disrupting Lifshitz transition can happen in the underdoped regime.\cite{Norman10}

The electron-electron (e-e) interaction may modify the single-particle picture. In particular, the e-e interaction in cuprates is believed to be strong, and may hold the key to understanding the mechanism of high-temperature superconductivity and other properties.\cite{Lee06} Here, we take the point of view in which the spin-density wave order is driven by the Coulomb interaction. Since the Lifshitz transitions in Figures \ref{fig:pd} and \ref{fig:nd} occur in states with well-established order (corresponding to relatively large density wave potentials), we assume a Fermi-liquid description for the electrons on the CuO$_2$ planes.\cite{Hackl10} 
In the framework of the Fermi liquid theory, the effect of e-e interaction on the Lifshitz transition has been studied.\cite{Kaganov84,Blanter94} In this paper, we concentrate on the effect of impurity scattering.

To lowest order in the electron-impurity interaction, the impurity scattering rate is proportional to the single-particle density of states. In three dimensions, the density of states acquires a $\sqrt{\pm(\mu-\epsilon_c)}$ non-analytic contribution in the vicinity of Lifshitz transitions, which has important effects on the thermodynamic and kinetic properties of the system.\cite{Varlamov89,Livanov99} The correction to the impurity scattering rate vanishes, albeit in a non-analytical way, as the transition is approached. The situation is different in two spatial dimensions, where for the pocket-disappearing type transition, the density of states has a finite jump, and for the neck-disrupting type, the density of states has a logarithmic singularity $\sim\ln|\mu-\epsilon_c|$.\cite{Blanter92} The logarithmically diverging density of states presents conceptual challenges; it was  pointed out that a system with diverging density of states at the Fermi energy is unstable to density wave ordering.\cite{Rice75} Furthermore, when the impurity scattering is large, perturbation theory cannot be applied. 

In this paper, we study Lifshitz transitions in two-dimensional systems, treating the impurity scattering in the self-consistent Born approximation, and calculate the transport coefficients in the low-temperature limit. We keep both real and imaginary parts of the electron self-energy in solving the self-consistent equation. For the neck-disrupting type transition, the divergence in the density of states is cut off by the impurity potential, and this approach leads to a conceptually consistent picture. Restricting to the low-temperature limit allows us to study the impurity effects and neglect the scattering from phonons and other collective excitations. The corrections to the self-consistent Born approximation consist of diagrams with crossed impurity lines, and as discussed in Refs [\onlinecite{Varlamov89,Blanter94}], they are negligible for the study of the electron self-energy. Their contribution to two-particle processes, such as the weak localization effect,\cite{Altshuler85} will not be considered. An external magnetic field suppresses the weak localization effect. A recent discussion on the localization effect from experimental point of view can be found in Ref [\onlinecite{LeBoeuf10}]. 

The rest of the paper is organized as follows. In Sec \ref{self energy}, we solve the electron self-energy in the self-consistent Born approximation. In Sec \ref{thermopower}, we study the low-temperature electrical conductivity and thermopower close to the Lifshitz transition. Sec \ref{summary} is a summary where we discuss the main results and their applications to underdoped cuprates. Some technical details are contained in the appendices. Throughout this paper, we take $\hbar=k_B=1$. 

\section{The self-consistent Born approximation for impurity scattering}
\label{self energy}

We use a mean-field approximation for the spin-density wave state in underdoped cuprates, in which electrons move on a square lattice subject to additional periodic potentials due to density wave ordering. This results in a multi-band situation with a band dispersion depending on density wave potentials.\cite{Lin05,Hackl10-2,Hackl10} Close to the transition, this dependence can be neglected, and we assume a rigid band model. The chemical potential is the tuning parameter for the transition.   

For impurity scattering, we assume point-like scatterers, such that the electron-impurity interaction matrix element is a constant in momentum space, denoted as $u$. As a result, the electron self-energy in the self-consistent Born approximation is independent of momentum,\cite{Mahan} 
\begin{equation}\label{eq:sigma}
\Sigma(i\epsilon_n)=u^2\int\frac{d^2p}{(2\pi)^2}\frac{1}{i\epsilon_n-\varepsilon_\mathbf{p}+\mu-\Sigma(i\epsilon_n)},
\end{equation}
where $\varepsilon_\mathbf{p}$ is the electron band dispersion and $\mu$ is the chemical potential. We study the situation where $\varepsilon_\mathbf{p}$ has only one extremal point at momentum $\mathbf{p}_L$ with energy $\varepsilon_c$ close to $\mu$, and assume that when several extrema are close to $\mu$, their effects are additive. We separate the momentum integral into a region close to $\mathbf{p}_L$, labeled by ``$L$" (the regions bounded by the dashed lines in Figures \ref{fig:pd} and \ref{fig:nd}), and a region far away from $\mathbf{p}_L$, labeled by ``$R$", and assume that there are extensive Fermi surface pieces in the $R$-region. For the momentum integral in the $R$-region, the usual arguments apply,\cite{AGD} and we obtain 
\begin{equation}\label{eq:gamma0}
u^2\int_R\frac{d^2p}{(2\pi)^2}\frac{1}{i\epsilon_n-\varepsilon_\mathbf{p}+\mu-\Sigma(i\epsilon_n)}\approx -i\gamma_0\mathrm{sign}\epsilon_n,
\end{equation}
where we have assumed, \textit{a posteriori}, that $|\Sigma(i\epsilon_n)+i\epsilon_n|\ll \mathcal{D}$ with $\mathcal{D}$ of the order of the bandwidth which is the largest energy scale in our calculation, and $\gamma_0=\pi u^2\bar{\nu}_0$ with $\bar{\nu}_0$ the density of electronic states in the $R$-region. $\gamma_0$ can be estimated from the scattering rate far from the Lifshitz transition. We assume weak impurity scattering, $\gamma_0\ll \mathcal{D}$. After analytical continuation $i\epsilon_n\to \epsilon+i0^+$, the retarded self-energy $\Sigma^R(\epsilon)$ can be written as $\Sigma^R(\epsilon)=-i\gamma_0+\delta\Sigma^R(\epsilon)$. Denoting  $R(\epsilon)$ and $-\gamma_1(\epsilon)$ as real and imaginary parts of $\delta\Sigma^R(\epsilon)$, we have 
\begin{equation*}
\Sigma^R(\epsilon)=R(\epsilon)-i\gamma_t(\epsilon),
\end{equation*}
where $\gamma_t(\epsilon)=\gamma_0+\gamma_1(\epsilon)$. From Eqs (\ref{eq:sigma}, \ref{eq:gamma0}), $R(\epsilon)$ and $\gamma_t(\epsilon)$ are determined from the integral equations 
\begin{align}
&R(\epsilon)=u^2\int_L\frac{d^2p}{(2\pi)^2}\frac{\epsilon+\mu-R(\epsilon)-\varepsilon_\mathbf{p}}{(\epsilon+\mu-R(\epsilon)-\varepsilon_\mathbf{p})^2+\gamma_t^2},\label{eq:R}
\\
&\gamma_t(\epsilon)=\gamma_0+u^2\int_L\frac{d^2p}{(2\pi)^2}\frac{\gamma_t(\epsilon)}{(\epsilon+\mu-R(\epsilon)-\varepsilon_\mathbf{p})^2+\gamma_t^2}.\label{eq:gammat}
\end{align}
In the low-temperature regime ($T\ll \gamma_t$), one can use the expansion $R(\epsilon)\approx R_0+R^\prime_0 \epsilon$ and $\gamma_t(\epsilon)\approx \gamma_{t0}+\gamma^\prime_{t0}\epsilon$. Further calculation requires a knowledge of $\varepsilon_\mathbf{p}$ in the $L$-region which is different for the two types of Lifshitz transitions, and is discussed in the following subsections.  

\subsection{The pocket-disappearing Lifshitz transition}

For the pocket-disappearing Lifshitz transition, the dispersion relation in the $L$-region  can be expanded as, taking $\mathbf{p}_L$ as the origin and $p_{x,y}$ along the principle axes, 
\begin{equation}\label{eq:ep pd}
\varepsilon_\mathbf{p}=\epsilon_c\pm\Bigl(\frac{p_x^2}{2m_x}+\frac{p_y^2}{2m_y}\Bigr),
\end{equation}
where we choose $\epsilon_c=0$ in the calculation. The upper sign corresponds to the emergence of an electron pocket for $\mu>0$ and the lower sign corresponds to that of a hole pocket for $\mu<0$. In this paper, we assume an electron pocket; the other case can be studied in the same way. The $L$-region is defined by $\sqrt{k_x^2+k_y^2}<\Lambda$, where $k_x=p_x/\sqrt{2m_x}$ and $k_y=p_y/\sqrt{2m_y}$. 

Assuming $\Lambda^2\gg \gamma_t$, Eqs (\ref{eq:R}, \ref{eq:gammat}) become 
\begin{align}
&R(\epsilon)=-U\ln\frac{\Lambda^2}{\sqrt{[\epsilon+\mu-R(\epsilon)]^2+\gamma_t(\epsilon)^2}}, \label{eq:R pd}\\
&\gamma_t(\epsilon)=\gamma_0+ U \Bigl\{\frac{\pi}{2}+\tan^{-1}\Bigl[\frac{\epsilon+\mu-R(\epsilon)}{\gamma_t(\epsilon)}\Bigr]\Bigr\}, \label{eq:ge pd}
\end{align}
where $U =u^2\sqrt{m_xm_y}/2\pi$ and is of the same order as $\gamma_0$. In the following discussion, we find that it is convenient to measure energies in units of $U$. 

\begin{figure}
\centering
\includegraphics[width=.5\textwidth]{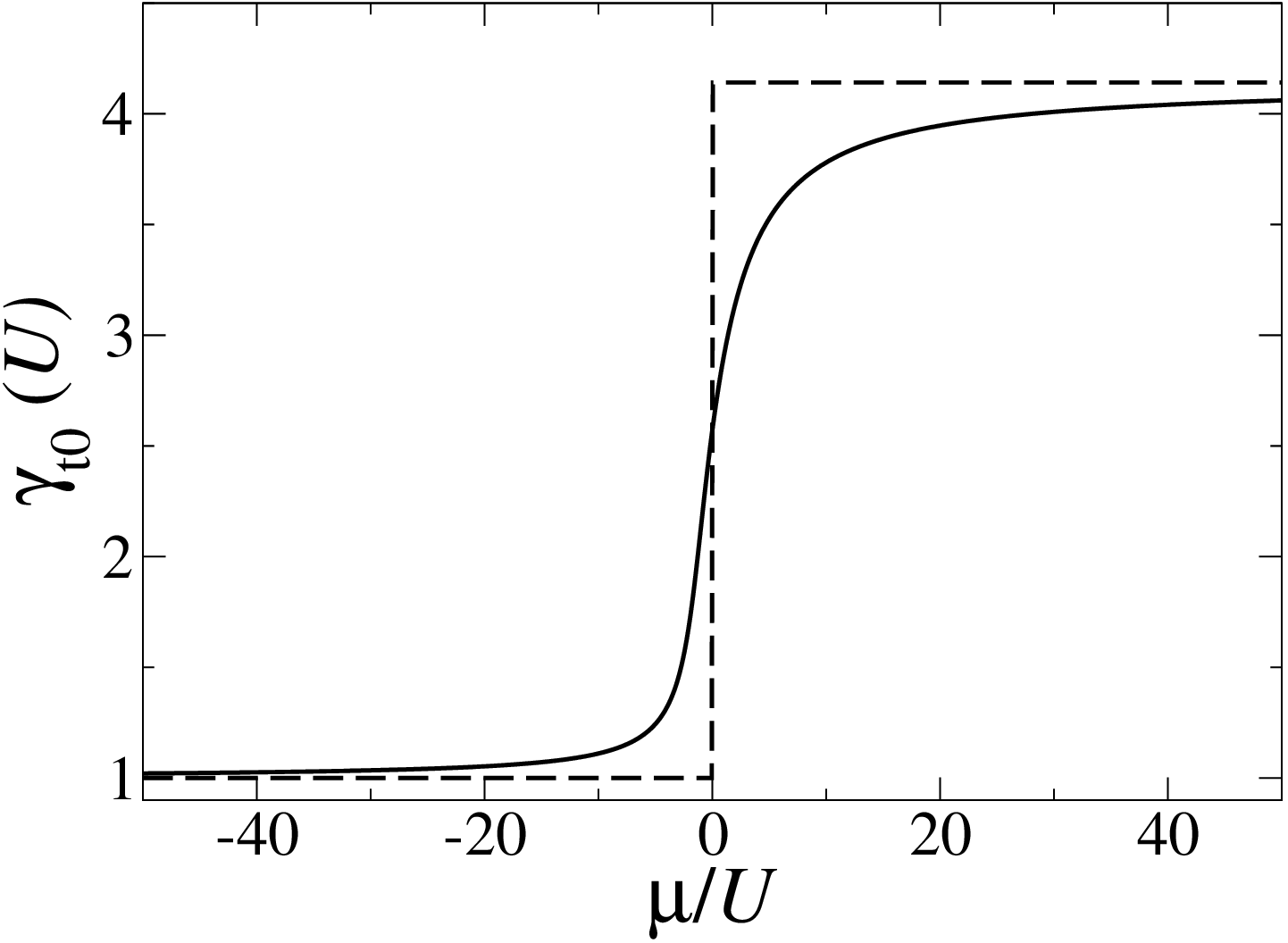}
\caption{\small{The impurity scattering rate $\gamma_{t0}$ close to the pocket-disappearing Lifshitz transition. Solid line: $\gamma_{t0}$ solved from the self-consistent Born approximation, Eq (\ref{eq:g0 pd}). Dashed line: $\gamma_{t0}^\mathrm{pt}=\gamma_0+ U \pi\Theta(\mu)$ in  perturbation theory. Here, $\gamma_0= U $.}}
\label{fig:gt pd}
\end{figure}

\begin{figure}
\centering
\includegraphics[width=.8\textwidth]{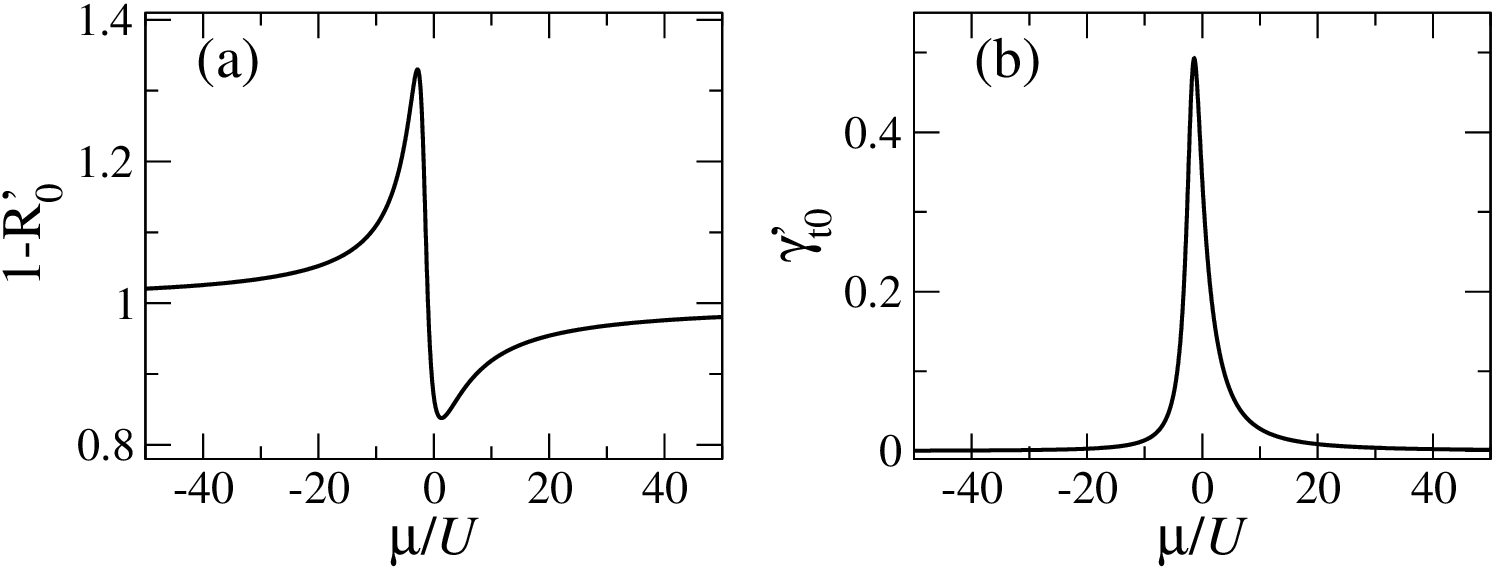}
\caption{\small{(a): $1-R_0^\prime$ and (b): $\gamma_{t0}^\prime$ close to the pocket-disappearing Lifshitz transition. Here $\gamma_0=U$.}}
\label{fig:zgp pd}
\end{figure}

From Eq (\ref{eq:R pd}), we obtain 
\begin{equation}\label{eq:R0 pd}
\mu=(\mu-R_0)-U\ln\frac{\Lambda^2}{\sqrt{(\mu-R_0)^2+\gamma_{t0}^2}},
\end{equation}
showing that $\mu$ is a monotonically increasing function of $\mu-R_0$. As can be seen from Eqs (\ref{eq:R pd}, \ref{eq:ge pd}), $\gamma_{t0}$, $\gamma_{t0}^\prime$, and $R_0$ all depend on $\mu-R_0$. As a result, we absorb $R_0$ into $\mu$. From Eq (\ref{eq:ge pd}), we obtain 
\begin{equation}\label{eq:g0 pd}
\gamma_{t0}=\gamma_0+ U \Bigl\{\frac{\pi}{2}+\tan^{-1}\frac{\mu}{\gamma_{t0}}\Bigr\},
\end{equation}
which is shown in Figure \ref{fig:gt pd}. $\gamma_{t0}^\prime$ and $1-R_0^\prime$ can be solved from Eqs (\ref{eq:R pd}, \ref{eq:ge pd}), and are shown in Figure \ref{fig:zgp pd}.

Since $|\tan^{-1}x|\leq \pi/2$, $\gamma_t(\epsilon)\sim  U $. For $\mu\gg  U $, $\gamma_{t0}\approx \gamma_0+\pi U $, $|R_0^\prime|\sim  U /\mu\ll 1$, and $|\gamma_{t0}^\prime|\sim  U ^2/\mu^2\ll 1$. In this limit, the electron pocket in the $L$-region is well established, with an energy scale large compared to impurity broadening, and this corresponds to the usual situation for impurity scattering where the imaginary part of $\Sigma^R(\epsilon)$ is proportional to the single-particle density of states, and the real part is negligible. For $-\mu\gg  U $, $\gamma_{t0}\approx\gamma_0$, $|R_0^\prime|\sim  U /|\mu|\ll 1$, and $|\gamma_{t0}^\prime|\sim  U ^2/\mu^2\ll 1$. In this limit, the states in the $L$-region are far below the chemical potential and are  inactive. For $|\mu|\lesssim  U $, $\gamma_{t0}$ is solved from Eq (\ref{eq:g0 pd}), $|R_0^\prime|\sim \mathcal{O}(1)$, and $|\gamma_{t0}^\prime|\sim \mathcal{O}(1)$. At the transition point ($\mu=0$), $\gamma_{t0}=\gamma_0+\frac{\pi}{2} U $, $R^\prime_0=1-(1+ U ^2/\gamma_{t0}^2)^{-1}$, and $\gamma_{t0}^\prime=\frac{ U }{\gamma_{t0}}(1+ U ^2/\gamma_{t0}^2)^{-1}$. Close to the Lifshitz transition, the real part of the self-energy is not negligible. 

In Figure \ref{fig:gt pd}, $\gamma_{t0}$ (solid line) is compared to the result from perturbation theory $\gamma_{t0}^\mathrm{pt}=\gamma_0+ U \pi\Theta(\mu)$ (dashed line), where $\Theta(x)=1$ for $x>0$ and 0 otherwise, which is obtained by substituting $\Sigma(i\epsilon_n)=i0^+\mathrm{sign}\epsilon_n$ into the integrand in Eq (\ref{eq:sigma}). There is a jump in $\gamma_{t0}^\mathrm{pt}$ at the transition, originating from that in the density of states when a new Fermi pocket appears in two-dimensional systems. From Figure \ref{fig:gt pd}, we note that for large $|\mu|$, $\gamma_{t0}\approx \gamma_{t0}^\mathrm{pt}$ and the self-consistency is not essential, while for small $|\mu|$, the jump in $\gamma_{t0}^\mathrm{pt}$ is replaced by a smooth crossover in $\gamma_{t0}$. Since the width of the crossover region and the jump in $\gamma_{t0}$ are both of order $ U $, we expect $\gamma_{t0}^\prime\sim \mathcal{O}(1)$ close to the Lifshitz transition ($|\mu|\lesssim  U $).

We note that in all cases $\gamma_t\sim  U $ and $1-R_0^\prime\sim 1$, and thus $|\Sigma(i\epsilon_n)+i\epsilon_n|\ll \mathcal{D}$. This justifies the calculation of the integral in Eq (\ref{eq:gamma0}).

\subsection{The neck-disrupting Lifshitz transition}

For the neck-disrupting Lifshitz transition, the dispersion relation $\varepsilon_\mathbf{p}$ in the $L$-region can be expanded as, choosing $\mathbf{p}_L$ as the origin and $p_{x,y}$ along the principle axes, 
\begin{equation}\label{eq:ep nd}
\varepsilon_\mathbf{p}\approx \epsilon_c+\frac{p_x^2}{2m_x}-\frac{p_y^2}{2m_y},
\end{equation}
where $\epsilon_c=0$. For convenience, the $L$-region is defined as $-\Lambda<k_x,k_y<\Lambda$ where $k_x=p_x/\sqrt{2m_x}$ and $k_y=p_y/\sqrt{2m_y}$. Using Appendix \ref{integrals}, Eqs (\ref{eq:R}, \ref{eq:gammat}) become 
\begin{align}
& R(\epsilon)=U\tan^{-1}\frac{\epsilon+\mu-R(\epsilon)}{\gamma_t(\epsilon)}, \label{eq:R nd}\\
&\gamma_t(\epsilon)=\gamma_0+ U \ln\frac{\Lambda^2}{\sqrt{\gamma_t(\epsilon)^2+[\epsilon+\mu-R(\epsilon)]^2}}\label{eq:ge nd}. 
\end{align}

\begin{figure}
\centering
\includegraphics[width=.5\textwidth]{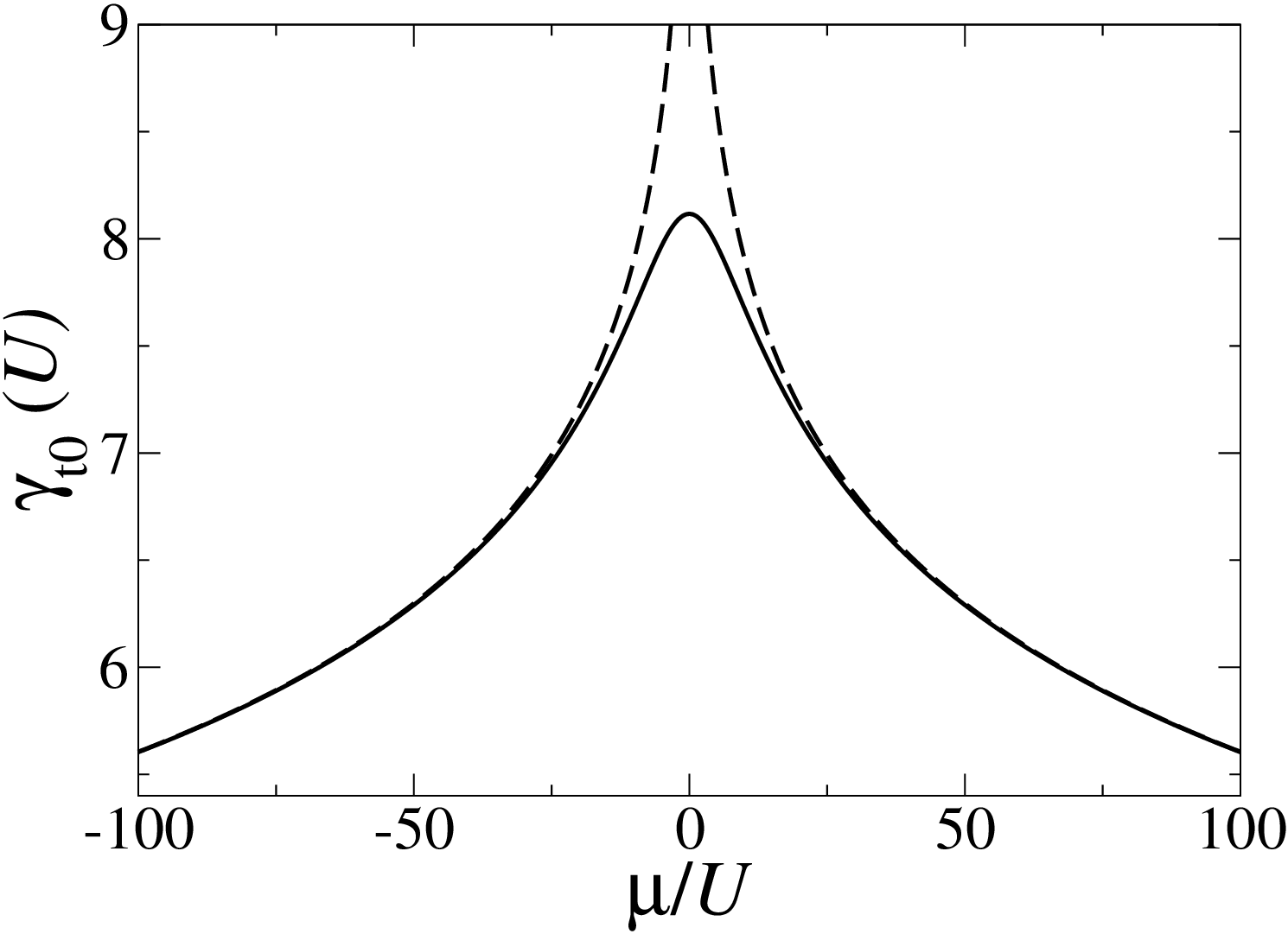}
\caption{\small{The impurity relaxation rate $\gamma_{t0}$ as a function of $\mu$ close to the neck-disrupting Lifshitz transition. Solid line: $\gamma_{t0}$ solved from the self-consistent Born approximation, Eq (\ref{eq:g0 nd}). Dashed line: $\gamma_{t0}^\mathrm{pt}=\gamma_0+ U \ln\frac{\Lambda^2}{|\mu|}$ from perturbation theory. Here, $\gamma_0=U$ and $\Lambda^2=10^4U$.}}
\label{fig:gt nd}
\end{figure}

From Eq (\ref{eq:R nd}), 
\begin{equation}
\mu=\mu-R_0+\tan^{-1}\frac{\mu-R_0}{\gamma_{t0}},
\end{equation}
which is again a monotonically increasing function. As discussed above, $R_0$ is absorbed into $\mu$. From Eq (\ref{eq:ge nd}), 
\begin{equation}\label{eq:g0 nd}
\gamma_{t0}=\gamma_0+ U \ln\frac{\Lambda^2}{\sqrt{\mu^2+\gamma_{t0}^2}},
\end{equation}
which is shown in Figure \ref{fig:gt nd}. We now discuss the role played by $\Lambda$, introduced for the expansion of $\varepsilon_\mathbf{p}$ in the vicinity of the band extrema in Eqs (\ref{eq:ep pd}, \ref{eq:ep nd}). One expects that $\Lambda^2\lesssim \mathcal{D}$. In the calculation, we assume that $U\ll\Lambda^2\ll\mathcal{D}$. Close to the pocket-disappearing transition, $\Lambda$ can be chosen in an obvious way, as suggested by Figure \ref{fig:pd}. Furthermore, $\Lambda$ only appears in Eq (\ref{eq:R0 pd}), which is absorbed into $\mu$. The neck-disrupting transition is more subtle; in Figure \ref{fig:nd}, the introduction of $\Lambda$ seems to be arbitrary, and furthermore $\Lambda$ appears explicitly in Eq (\ref{eq:ge nd}). When $\Lambda^2\ll \mathcal{D}$ such that Eq (\ref{eq:ep nd}) is also applicable in part of the $R$-region, $\gamma_0$ also depends on $\Lambda$, and this dependence cancels the $\Lambda$-dependence in $\gamma_1$. Thus, $\gamma_t$ is independent of the choice of $\Lambda$.

\begin{figure}
\centering
\includegraphics[width=.8\textwidth]{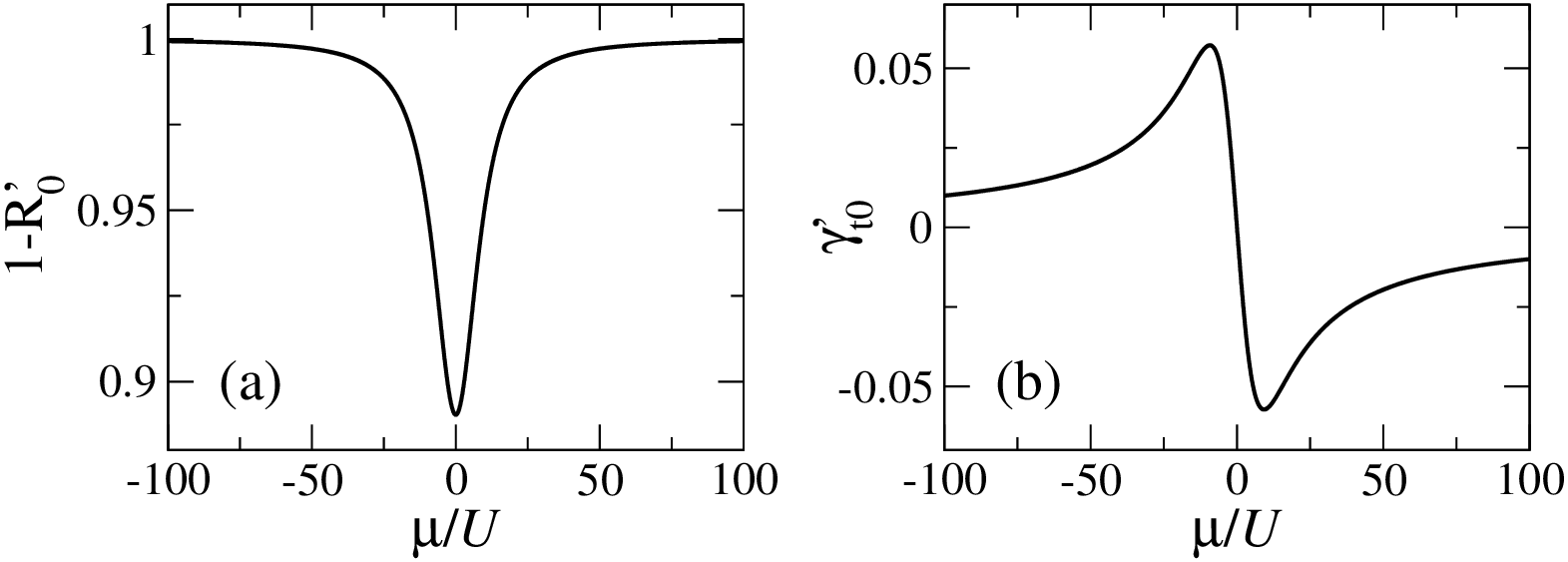}
\caption{\small{(a): $1-R_0^\prime$ and (b): $\gamma_{t0}^\prime$ close to the neck-disrupting Lifshitz transition. Here, $\gamma_0=U$ and $\Lambda=10^4U$. }}
\label{fig:zgp nd}
\end{figure}

In Eq (\ref{eq:g0 nd}), for $|\mu|\gg \gamma_{t0}$, $\gamma_{t0}\approx \gamma_0+ U \ln\frac{\Lambda^2}{|\mu|}$. In this limit, $R_0^\prime\sim \frac{ U ^2}{\mu^2}\ln\frac{\Lambda^2}{|\mu|}\ll 1$, and $|\gamma_{t0}^\prime|\sim  U /|\mu|\ll 1$. On the other hand, when $|\mu|\sim \gamma_{t0}\approx  U \ln\frac{\Lambda^2}{ U }$, $|R_0^\prime|\sim 1/\ln\frac{\Lambda^2}{ U }$ and $|\gamma_{t0}^\prime|\sim 1/\ln\frac{\Lambda^2}{ U }$. For $\mu=0$, $\gamma_{t0}^\prime=0$. In Figure \ref{fig:gt nd}, $\gamma_{t0}$ (solid line) is compared to $\gamma_{t0}^\mathrm{pt}=\gamma_0+ U \ln\frac{\Lambda^2}{|\mu|}$ from perturbation theory (dashed line).\cite{Blanter92} The logarithmic divergence in $\gamma_{t0}^\mathrm{pt}$ is replaced by a finite peak with height $\sim  U \ln\frac{\Lambda^2}{ U }$. $1-R_0^\prime$ and $\gamma_{t0}^\prime$ are shown in Figure \ref{fig:zgp nd}, from which we see that $R_0^\prime$ and $\gamma_{t0}^\prime$ are small ($\lesssim 0.1$) in the entire range plotted. However, because $\ln x$ increases rather slowly for $x\gg 1$, they are not negligible at $|\mu|\sim  U \ln\frac{\Lambda^2}{ U }$ even for $\Lambda^2=10^4 U $. 

Since $\gamma_{t0}\lesssim  U \ln\frac{\Lambda^2}{ U }$ and $1-R_0^\prime\sim 1$, $|\Sigma(i\epsilon_n)+i\epsilon_n|\ll \mathcal{D}$ is satisfied for a weak impurity potential $ U \ll \Lambda^2$. This justifies the evaluation of the momentum integral in Eq (\ref{eq:gamma0}).

\section{The low-temperature electrical conductivity and thermopower close to the Lifshitz transition}
\label{thermopower}

In order to study the thermopower, defined in Eq (\ref{eq:thermopower}), close to the Lifshitz transition, we first calculate the electrical conductivity $\sigma_{xx}$ and the thermoelectric transport coefficient $\alpha_{xx}$. We separate the momentum integrals in Eqs (\ref{eq:sigmaab}, \ref{eq:alphaab}) into $R$- and $L$-regions, and introduce  
\begin{equation}\label{eq:RL}
\begin{bmatrix}\sigma_{xx}^{R,L}\\ \alpha_{xx}^{R,L}\end{bmatrix}=\frac{e}{2\pi}\int_{-\infty}^\infty\frac{dx}{\cosh^2(x/2)}\begin{bmatrix} e \\ x\end{bmatrix} \int_{R,L}\frac{d^2p}{(2\pi)^2}\bigl(v_p^x\bigr)^2\bigl\{\mathrm{Im}G^R(\mathbf{p},xT)\bigr\}^2,
\end{equation}
where $G^R(\mathbf{p},\epsilon)=\frac{1}{(1-R_0^\prime)\epsilon-\varepsilon_\mathbf{p}+\mu+i\gamma_t(\epsilon)}$. Then, $\sigma_{xx}=\sigma_{xx}^R+\sigma_{xx}^L$ and $\alpha_{xx}=\alpha_{xx}^R+\alpha_{xx}^L$. 

In the $R$-region, we define the function $\mathcal{F}_R(\xi)$ as 
\begin{equation}\label{eq:F}
\mathcal{F}_R(\xi)=\int_R\frac{d^2p}{(2\pi)^2}\delta (\varepsilon_\mathbf{p}-\mu-\xi)\bigl(v_p^x\bigr)^2,
\end{equation}
in terms of which, $\sigma_{xx}^R$ and $\alpha_{xx}^R$ can be written as 
\begin{equation}\label{eq:R formula}
\begin{bmatrix}\sigma_{xx}^R\\ \alpha_{xx}^R\end{bmatrix}=\frac{e}{2\pi}\int_{-\infty}^\infty\frac{dx}{\cosh^2(x/2)}\begin{bmatrix}e \\ x\end{bmatrix} \int d\xi \mathcal{F}_R(\xi)\Bigl\{\frac{\gamma_t(xT)}{(\xi-(1-R_0^\prime)xT)^2+\gamma_t(xT)^2}\Bigr\}^2.
\end{equation}
The integral is dominated by $x\lesssim 1$, and for $T\ll \gamma_t$, in order to obtain a non-zero $\alpha_{xx}^R$, we expand the integrand to order $xT$. One contribution comes from the expansion $\mathcal{F}_R(\xi)\approx \mathcal{F}_{R0}+\mathcal{F}_{R0}^\prime \xi$, where $\mathcal{F}_{R0}^\prime\lesssim \mathcal{F}_{R0}/\Lambda^2$ and is negligible. By neglecting contributions of order $1/\Lambda^2$, we are calculating the singular contribution to $\alpha_{xx}$ due to the Lifshitz transition. Thus, 
\begin{equation}\label{eq:R result}
\begin{bmatrix} \sigma_{xx}^R\\ \alpha_{xx}^R\end{bmatrix}=\frac{e}{4}\mathcal{F}_{R0}\int_{-\infty}^\infty \frac{dx}{\cosh^2(x/2)}\begin{bmatrix} e\\ x\end{bmatrix}\frac{1}{\gamma_t(xT)}. 
\end{equation}

In Appendix \ref{L region}, we argue that, close to the Lifshitz transition, $\sigma_{xx}^L$ and $\alpha_{xx}^L$ are both negligible. As a result, $\sigma_{xx}$ and $\alpha_{xx}$ are given by Eq (\ref{eq:R result}). Performing the remaining integration, 
\begin{equation}\label{eq:sxx result}
\sigma_{xx}\approx e^2\mathcal{F}_{R0}/\gamma_{t0},
\end{equation}
and 
\begin{equation}\label{eq:axx result}
\alpha_{xx}\approx -\frac{\pi^2}{3}eT\mathcal{F}_{R0}\frac{\gamma_{t0}^\prime}{\gamma_{t0}^2}.
\end{equation}
Eqs (\ref{eq:sxx result}, \ref{eq:axx result}) can be used for both types of Lifshitz transitions; both $\sigma_{xx}$ and $\alpha_{xx}$ are determined by states in the $R$-region, and the effect of the transition is manifested through $\gamma_{t}(\epsilon)$. 

The resistivity $\rho_{xx}=1/\sigma_{xx}\propto \gamma_{t0}$. Thus, up to scale factors, $\rho_{xx}$ is given by the solid curve in Figure \ref{fig:gt pd} for the pocket-disappearing transition and the solid curve in Figure \ref{fig:gt nd} for the neck-disrupting transition. 

We now give an estimate of $\sigma_{xx}^N$ and $\alpha_{xx}^N$ for a normal metallic state, which  are obtained from Eq (\ref{eq:R formula}) with $\mathcal{F}_R(\xi)$ replaced by $\mathcal{F}(\xi)$ which is defined by Eq (\ref{eq:F}) with a momentum integration over the entire Brillouin zone. We note that $\mathcal{F}\sim \mathcal{F}_R$. Since $d\mathcal{F}(\xi)/d\xi\sim \mathcal{F}/\mathcal{D}$ and $d\gamma/d\epsilon\sim \gamma/\mathcal{D}$ with $\gamma\sim  U $, $\sigma_{xx}^N\sim e^2\mathcal{F}/ U $ and $|\alpha_{xx}^N|\sim \frac{\pi^2}{3}eT\frac{\mathcal{F}}{ U }\frac{1}{\mathcal{D}}$. The thermopower of a normal metallic state is then $|Q_N|=|\alpha^N_{xx}|/\sigma_{xx}^N\sim \frac{\pi^2}{3e}T/\mathcal{D}$, which is usually small.\cite{Behnia09}

\begin{figure}
\centering
\includegraphics[width=.8\textwidth]{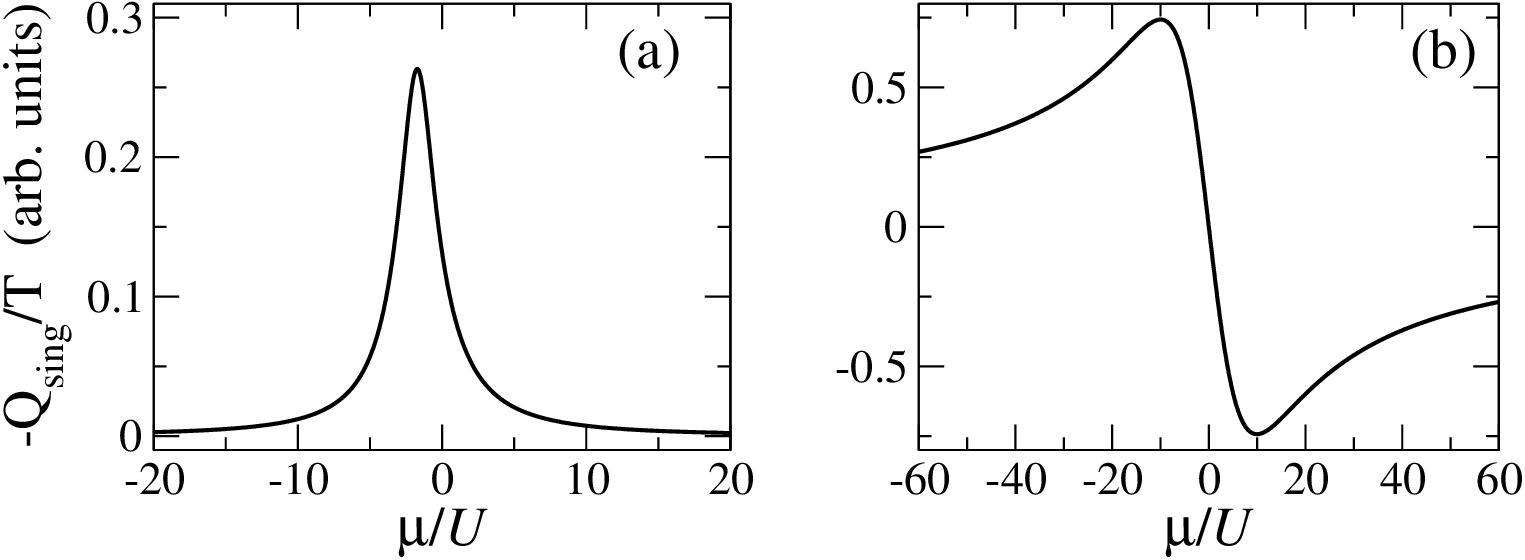}
\caption{\small{The singular part of the low-temperature thermopower, $-Q_\mathrm{sing}/T$, for the pocket-disappearing  Lifshitz transition (a) and for the neck-disrupting Lifshitz transition (b). $\gamma_0=U$ and $\Lambda^2=10^4U$. }}
\label{fig:Q}
\end{figure}

Close to Lifshitz transitions, the singular contribution to the thermopower $Q$ is 
\begin{equation}\label{eq:Q result}
Q_\mathrm{sing}=\frac{\alpha_{xx}}{\sigma_{xx}}\approx -\frac{\pi^2}{3e}T\frac{\gamma_{t0}^\prime}{\gamma_{t0}}=-\frac{\pi^2}{3e}T\Bigl[\frac{d\ln\gamma_t(\epsilon)}{d\epsilon}\Bigr]_{\epsilon=0}.
\end{equation}
Figure \ref{fig:Q} (a) shows $-Q_\mathrm{sing}/T$ close to the pocket-disappearing transition,  which is characterized by a sharp peak for $|\mu|\lesssim  U $,  arising from the sharp peak in $\gamma_{t0}^\prime$ (Figure \ref{fig:zgp pd} (b)).  As discussed in Sec \ref{self energy}, close to the transition, $\gamma_{t0}^\prime\sim \mathcal{O}$(1). As a consequence, $|Q_\mathrm{sing}/Q_N|\sim \mathcal{D}/ U \gg 1$, suggesting a strong enhancement over the normal state value for $|\mu|\lesssim  U $. Figure \ref{fig:Q} (b) shows $-Q_\mathrm{sing}/T$ close to the neck-disrupting Lifshitz transition, from which we see that $Q_\mathrm{sing}$ changes sign at the transition, and $|Q_\mathrm{sing}|$ has two peaks at $\mu\sim \pm U \ln\frac{\Lambda^2}{ U }$. For $|\mu|\sim \gamma_{t0}$, $|Q_\mathrm{sing}/Q_N|\sim \frac{\mathcal{D}/ U }{(\ln\frac{\Lambda^2}{ U })^2}\gg 1$, and for $|\mu|\gg \gamma_{t0}$, $|Q_\mathrm{sing}/Q_N|\sim \frac{\mathcal{D}/|\mu|}{\ln\frac{\Lambda^2}{|\mu|}}\gg 1$. Thus, close to the neck-disrupting transition, the thermopower is strongly enhanced compared to the normal state value. 

We now give a qualitative discussion on relating $\mu$ to the doping $x$ which is usually easier to control experimentally. Assuming a rigid-band model (the density wave potential is fixed), close to $x_L$ (the critical doping), 
\begin{equation}\label{eq:mu}
x-x_L\approx \mathcal{A} \mu/t_1,
\end{equation}
where $t_1$ is the leading hopping integral for the cuprate band dispersion.\cite{Andersen95,Lin05} In Eq (\ref{eq:mu}), we have neglected complications due to Lifshitz transitions; in the pocket-disappearing type, $\mathcal{A}_+$ for $x>x_L$ is different from $\mathcal{A}_-$ for $x<x_L$ with $\mathcal{A}_+/\mathcal{A}_-\sim \mathcal{O}(1)$, and in the neck-disrupting type, $\mathcal{A}$ has a logarithmic divergence as $\mu\to 0$, which is cut off by the impurity scattering. For a qualitative estimation, we assume Eq (\ref{eq:mu}), and thus obtain 
\begin{equation}\label{eq:x0}
\mu/U=(x-x_L)/x_0,
\end{equation}
where $x_0\approx\mathcal{A}U/t_1\lesssim\gamma_0/t_1$, where $\mathcal{A}\lesssim 1$ since the total band width is $>t_1$. 

\section{Summary and discussion}
\label{summary}

In this paper, we have discussed the effects of impurity scattering on Lifshitz transitions in two-dimensional systems, with applications to underdoped cuprates. We solved the electron self-energy from the self-consistent Born approximation, and studied the low-temperature electrical conductivity and thermopower. We find that the conductivity and the thermopower are determined by states in the $R$-region, away from the momentum where the Fermi surface changes at the transition. The effects of the transition are manifested by the impurity scattering relaxation rate and its derivative. Close to the transition, the electrical resistivity is proportional to the relaxation rate $\gamma_{t0}$, and the thermopower is proportional to the derivative of the logarithm of the relaxation rate and is strongly enhanced compared to that of a normal metal, $|Q/Q_N|\sim t_1/\gamma_0$ where $Q_N$ and $\gamma_0$ are, respectively, the thermopower and the relaxation rate away from the transition, and $t_1$ is the leading hopping integral of the cuprate band structure.\cite{Andersen95}

Recent thermopower measurements on both electron-doped cuprates\cite{Li07} and hole-doped cuprates (La$_{1.6-x}$Nd$_{0.4}$Sr$_x$CuO$_4$\cite{Daou09} and YBa$_2$Cu$_3$O$_{6.67}$ in a magnetic field\cite{Chang10}) have provided evidence for spin-density wave ground states and quantum phase transitions into such states. In this paper, we see that by tuning the system through Lifshitz transitions, via \textit{e.g.} doping, the thermopower shows characteristic behavior. This provides another perspective on the spin-density wave state in underdoped cuprates. For electron-doped cuprates, a commensurate $(\pi,\pi)$ spin-density wave state has been proposed as the ground state when the superconductivity is suppressed.\cite{Armitage09} In this scenario, the doping level at which the hole pocket disappears and a pocket-disappearing Lifshitz transition occurs depends on how fast the spin potential grows with underdoping. Hall effect calculations suggest that this happens at a doping $x_L\approx 0.12$.\cite{Lin05} In this paper, we argue that the signatures in the thermopower (Figure \ref{fig:Q} (a)) can be used to determine the value of $x_L$. For the hole-doped materials YBa$_2$Cu$_3$O$_{6+x}$, the Fermi surface reconstruction due to the stripe order is more complicated.\cite{Lin08,Hackl10-2} Recent Hall effect measurement suggests a pocket-disappearing transition,\cite{LeBoeuf10} and the disappearance of quantum oscillations has been interpreted as due to a neck-disrupting transition.\cite{Norman10} Our results suggest that a measurement of the low-temperature thermopower may provide more information about this material. 

One advantage of our proposal is that Lifshitz transitions take place deep in the ordered states, and thus quantum fluctuations are expected to be not important. The potential disadvantage is that, as shown in Figure \ref{fig:Q}, the signatures in the thermopower are prominent only in the close vicinity of the transition. Eq (\ref{eq:x0}) suggests that the doping $x$ is varied at the scale of $x_0\lesssim \gamma_0/t_1$, which may present challenges in experimental studies.  

We note that our approach also applies to other density-wave orders proposed for the underdoped cuprates, \textit{e.g.} the d-density wave order.\cite{Chakravarty01,Dimov08} It can also be extended to study other two-dimensional systems close to Lifshitz transitions.

\textit{Acknowledgement} The author thanks Alex Levchenko for discussions, and A. J. Millis and M. R. Norman for advice and reading of the manuscript. This work was supported by the US DOE, Office of Science, under contract DE-AC02-06CH11357 and by the Center for Emergent Superconductivity, an Energy Frontier Research Center funded by the US DOE, Office of Science, under Award No.~DE-AC0298CH1088.

\appendix

\section{the integrals in the neck-disrupting Lifshitz transition}
\label{integrals}

In this Appendix, we present a general method to calculate integrals of the form 
\begin{equation}
I[F]=\int_{-\Lambda}^\Lambda \int_{-\Lambda}^\Lambda dk_x dk_y f(k_x^2-k_y^2)F(k_x,k_y),
\end{equation}
which appears in the discussion of the neck-disrupting Lifshitz transition. As usual, we make the transformation $\int dk_x dk_y\to \int d\xi dk_t\frac{1}{v}$, where $\xi=k_x^2-k_y^2$ which can be viewed as the energy variable, $v=2\sqrt{k_x^2+k_y^2}$, and $dk_t^2=dk_x^2+dk_y^2$ is the line element along the constant energy contour. 

\begin{figure}
\centering
\includegraphics[width=.6\textwidth]{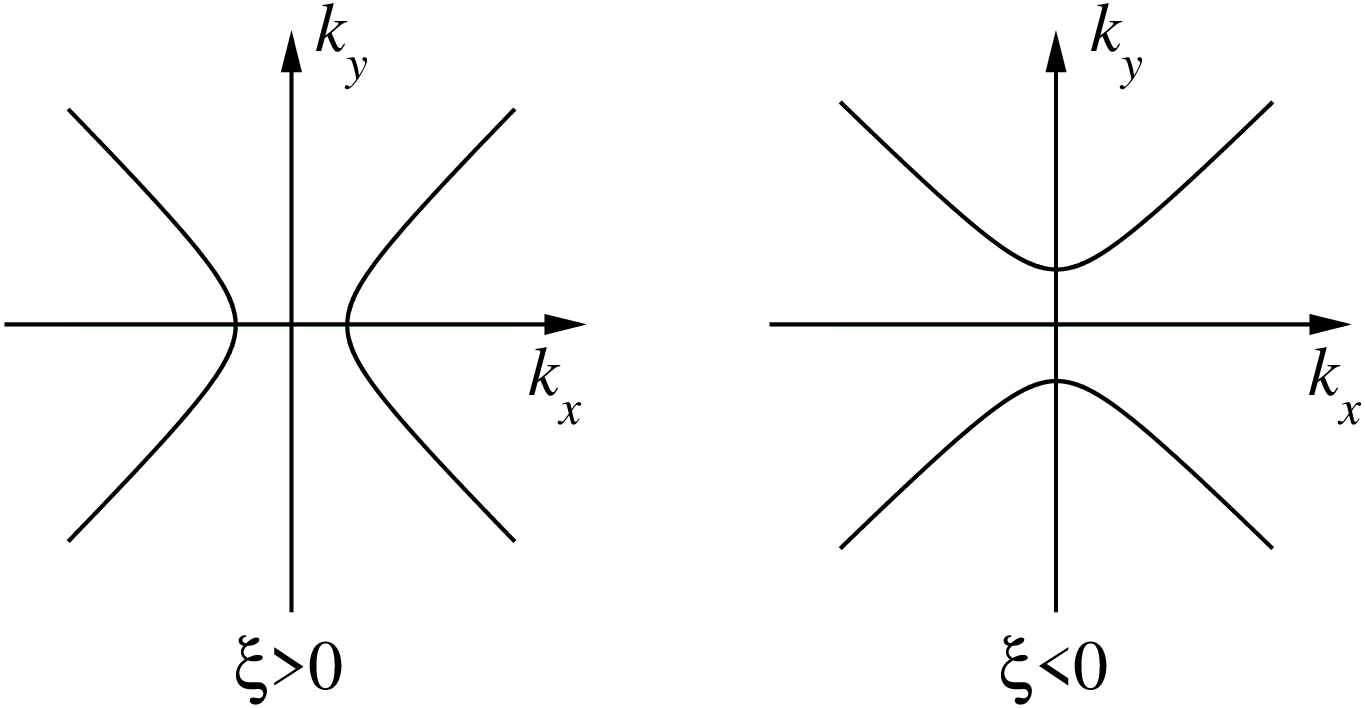}
\caption{\small{The constant energy contour for $k_x^2-k_y^2=\xi$.}}
\label{fig:contour}
\end{figure}
For $\xi>0$, the constant energy contour is $k_x=\pm\sqrt{\xi+k_y^2}$ with $-\Lambda\leq k_y\leq \Lambda$, while for $\xi<0$, the constant energy contour is $k_y=\pm\sqrt{k_x^2+|\xi|}$ with $-\Lambda\leq k_x\leq \Lambda$, as plotted in Figure \ref{fig:contour}. Thus, 
\begin{equation}
\begin{split}
I[F]=&\frac{1}{2}\int_{\xi>0}d\xi\int_{-\Lambda}^\Lambda dk_y\frac{f(\xi)}{\sqrt{|\xi|+k_y^2}}\Bigl(F(\sqrt{k_y^2+|\xi|},k_y)+F(-\sqrt{k_y^2+|\xi|},k_y)\Bigr)\\
&+\frac{1}{2}\int_{\xi<0}d\xi\int_{-\Lambda}^\Lambda dk_x \frac{f(\xi)}{\sqrt{|\xi|+k_x^2}}\Bigl(F(k_x,\sqrt{k_x^2+|\xi|})+F(k_x,-\sqrt{k_x^2+|\xi|})\Bigr).
\end{split}
\end{equation}

As an example, we consider the case $F=1$ and $f(\xi)=\delta(\xi-\epsilon)$, where $I[F]$ is proportional to  the single-particle density of states, 
\begin{equation}\label{eq:dos}
I[F]=\int d\xi \int_{-\Lambda}^\Lambda dk \frac{1}{\sqrt{|\xi|+k^2}}\delta (\xi-\epsilon)\approx2 \ln\frac{\Lambda}{\sqrt{|\epsilon|}}, 
\end{equation}
for $\Lambda\gg \sqrt{|\epsilon|}$.

\section{Transport coefficients in linear response theory}
\label{transport}

In this Appendix, we define the conductivity and the thermoelectric coefficients, and discuss their calculations in linear response theory. In the linear response regime, the electric current $\mathbf{J}$  flowing in response to an applied external electric field $\mathbf{E}$ and/or temperature gradient $\nabla T$ is given by 
\begin{equation}\label{eq:lr}
J_a=\sigma_{ab}E_b-\alpha_{ab}\nabla_b T, 
\end{equation}
where $a,b=x,y$ and repeated indices represent summation. 

Using the Kubo formula,\cite{Larkin} 
\begin{equation}
\sigma_{ab}=-\lim_{\omega\to 0}\frac{\mathrm{Im}[\mathcal{Q}^{ab}_{ee}(\omega)]}{\omega},\qquad \alpha_{ab}=-\frac{1}{T}\lim_{\omega\to 0}\frac{\mathrm{Im}[\mathcal{Q}^{ab}_{eq}(\omega)]}{\omega},
\end{equation}
where $\mathcal{Q}^{ab}_{ee,eq}(\omega)$ is obtained from $\mathcal{Q}_{ee,eq}^{ab}(i\omega_\nu)$ by analytical continuation $i\omega_\nu\to \omega+i0^+$, with 
\begin{equation}\label{eq:Qee}
\mathcal{Q}_{ee}^{ab}(i\omega_\nu)=2e^2T\sum_{i\epsilon_n}\int \frac{d^2p}{(2\pi)^2}v_p^av_p^b G(\mathbf{p},i\epsilon_n) G(\mathbf{p},i\epsilon_n+i\omega_\nu),
\end{equation}
and 
\begin{equation}\label{eq:QeQ}
\mathcal{Q}_{eq}^{ab}(i\omega_\nu)=2eT\sum_{i\epsilon_n}\int\frac{d^2p}{(2\pi)^2} (i\epsilon_n+i\omega_\nu/2)v_p^av_p^b G(\mathbf{p},i\epsilon_n)G(\mathbf{p},i\epsilon_n+i\omega_\nu).
\end{equation}
Note that in the present approximation, the self-energy $\Sigma(i\epsilon_n)$ has no momentum-dependence. As a result, we have used bare electric and heat current vertices in Eqs (\ref{eq:Qee}, \ref{eq:QeQ}). The frequency summation and the analytical continuation are standard.\cite{Mahan} The results are 
\begin{equation}\label{eq:sigmaab}
\sigma_{ab} =\frac{e^2}{2\pi}\int_{-\infty}^\infty \frac{dx}{\cosh^2(x/2)}\int \frac{d^2p}{(2\pi)^2}v_p^av_p^b \bigl\{\mathrm{Im}G^R(\mathbf{p},xT)\bigr\}^2,
\end{equation}
and 
\begin{equation}\label{eq:alphaab}
\alpha_{ab}=\frac{e}{2\pi}\int_{-\infty}^\infty\frac{xdx}{\cosh^2(x/2)}\int\frac{d^2p}{(2\pi)^2}v_p^av_p^b \bigl\{\mathrm{Im}G^R(\mathbf{p},xT)\bigr\}^2.
\end{equation}
Eqs (\ref{eq:sigmaab}, \ref{eq:alphaab}) are derived for a single-band system. It applies also to the multi-band situation (\textit{e.g.} the spin-density wave ordered state) when the impurity-scattering induced interband transition can be neglected. 

In the quasi-particle approximation,\cite{Mahan} $\bigl\{\mathrm{Im}G^R(\mathbf{p},\epsilon)\bigr\}^2\approx \pi\delta(\epsilon-\varepsilon_\mathbf{p}+\mu)/2\gamma$, Eqs (\ref{eq:sigmaab}, \ref{eq:alphaab}) lead to the familiar results for $\sigma_{ab}$ and $\alpha_{ab}$ obtained from the Boltzmann equation in the relaxation time approximation where the relaxation time $\tau\approx 1/2\gamma$. This gives a justification for the electric and heat current operators we have used. 

The thermopower $Q$ is defined as the coefficient of the electric field $E$ generated by an applied temperature gradient in the absence of the electric current. From Eq (\ref{eq:lr}), 
\begin{equation}\label{eq:thermopower}
Q=\frac{E_x}{\nabla_x T}=\frac{\alpha_{xx}}{\sigma_{xx}}. 
\end{equation}

\section{The calculation of $\sigma_{xx}^L$ and $\alpha_{xx}^L$}
\label{L region}

In this Appendix, we give a detailed discussion on the smallness of $\sigma_{xx}^L$ and $\alpha_{xx}^L$. The calculation is different for the two types of Lifshitz transitions, and is discussed in the following two subsections.

\subsection{The pocket-disappearing Lifshitz transition}

For the pocket-disappearing type with $\varepsilon_\mathbf{p}$ in the $L$-region given by Eq (\ref{eq:ep pd}), after rescaling momenta $p_{x,y}=\sqrt{2m_{x,y}}k_{x,y}$, $\sigma_{xx}^L$ and $\alpha_{xx}^L$ in Eq (\ref{eq:RL}) become 
\begin{equation*}
\begin{bmatrix}\sigma_{xx}^L\\\alpha_{xx}^L\end{bmatrix}=\frac{e}{2\pi}\int_{-\infty}^\infty \frac{dx}{\cosh^2(x/2)}\begin{bmatrix} e \\ x\end{bmatrix} \frac{1}{\pi^2}\sqrt{\frac{m_y}{m_x}}\int_{|\mathbf{k}|<\Lambda}d^2k\frac{k_x^2\gamma_{t}^2}{\bigl[(k^2-(1-R_0^\prime)xT-\mu)^2+\gamma_{t}^2\bigr]^2}.
\end{equation*}

We first consider $\sigma_{xx}^L$. In the limit $T\ll \gamma_{t0}$, 
\begin{equation*}
\sigma_{xx}^L\approx \frac{e^2}{2\pi^2}\sqrt{\frac{m_y}{m_x}}\Bigl\{1+\Bigl(\frac{\pi}{2}+\tan^{-1}\frac{\mu}{\gamma_{t0}}\Bigr)\frac{\mu}{\gamma_{t0}}\Bigr\},
\end{equation*}
with corrections $\sim T^2$. For both $-\mu\gg \gamma_{t0}$ and $|\mu|\lesssim \gamma_{t0}$, $\sigma_{xx}^L\sim e^2\sqrt{m_y/m_x}$. In these two cases, $\sigma_{xx}^L\ll \sigma_{xx}^R$ since we expect that $\mathcal{F}_{R0}/\gamma_{t0}$ is large for a weak impurity scattering potential. For $\mu\gg \gamma_{t0}$, $\sigma_{xx}^L\approx \frac{e^2}{2\pi}\sqrt{\frac{m_y}{m_x}}\frac{\mu}{\gamma_{t0}}$. In this limit, $\sigma_{xx}^L$ gives a small correction to $\sigma_{xx}^R$ since $\mathcal{F}_{R0}\sim \mathcal{D} \gg \mu$ close to the transition. We thus argue that $\sigma_{xx}^L$ is always negligible compared to $\sigma_{xx}^R$.  

For $\alpha_{xx}^L$, we find 
\begin{equation*}
\alpha_{xx}^L\approx\frac{\pi T}{6}\sqrt{\frac{m_y}{m_x}}\Bigl(\frac{\pi}{2}+\tan^{-1}\frac{\mu}{\gamma_{t0}}+\frac{\gamma_{t0}\mu}{\gamma_{t0}^2+\mu^2}\Bigr)\Bigl(\frac{1-R_0^\prime}{\gamma_{t0}}-\frac{\mu}{\gamma_{t0}^2}\gamma_{t0}^\prime\Bigr).
\end{equation*}
For $|\mu|\sim \gamma_{t0}$, $\alpha_{xx}^L\sim 1/\gamma_{t0}$ and $\alpha_{xx}^R\sim \mathcal{F}_{R0}/\gamma_{t0}^2$, and thus $|\alpha_{xx}^L|\ll |\alpha_{xx}^R|$. For $-\mu\gg\gamma_{t0}$, $|\alpha_{xx}^L|\ll|\alpha_{xx}^R|$ since the states in the $L$-region are far below the Fermi energy and are not contributing to the transport properties. For $\mu\gg\gamma_{t0}$, $\alpha_{xx}^L\sim \frac{\mu}{\gamma_{t0}^2}\gamma_{t0}^\prime$, and $|\alpha_{xx}^L|\ll |\alpha_{xx}^R|$ since $|\mu|\ll \mathcal{F}_{R0}$. Thus, $\alpha_{xx}^L$ is negligible compared to $\alpha_{xx}^R$.

\subsection{The neck-disrupting Lifshitz transition}

For the neck-disrupting type with $\varepsilon_\mathbf{p}$ in the $L$-region given by Eq (\ref{eq:ep nd}), we calculate the momentum integral using the method outlined in Appendix \ref{integrals},
\begin{equation*}
\begin{bmatrix}\sigma_{xx}^L\\\alpha_{xx}^L\end{bmatrix}=\frac{e}{2\pi}\int_{-\infty}^\infty\frac{dx}{\cosh^2(x/2)}\begin{bmatrix} e\\ x \end{bmatrix} \frac{2}{\pi^2}\sqrt{\frac{m_y}{m_x}}\int d\xi \frac{\gamma_t^2\bigl(\Lambda^2+\xi\ln\frac{\Lambda}{\sqrt{|\xi|}}\bigr)}{\bigl[(\xi-(1-R_0^\prime)xT-\mu)^2+\gamma_t^2\bigr]^2}. 
\end{equation*}

For $\sigma_{xx}^L$, 
\begin{equation*}
\sigma_{xx}^L\approx \frac{2e^2}{\pi^2}\sqrt{\frac{m_y}{m_x}}\frac{\Lambda^2}{\gamma_{t0}}+\frac{e^2}{\pi^2}\sqrt{\frac{m_y}{m_x}}\frac{\mu}{\gamma_{t0}}\ln\frac{\Lambda^2}{\sqrt{\mu^2+\gamma_{t0}^2}},
\end{equation*}
with corrections of order $T^2$. Compared to $\sigma_{xx}^R$, the first term is smaller by a factor $\sim \Lambda^2/\mathcal{D}\ll 1$. Since $\gamma_{t0},|\mu|\ll\Lambda^2$, the second term is smaller than the first one. Thus, $\sigma_{xx}^L$ can be neglected. 

For $\alpha_{xx}^L$, to leading order in $T$, we obtain 
\begin{equation*}
\alpha_{xx}^L\approx \frac{1}{3}eT\sqrt{\frac{m_y}{m_x}} \Bigl\{\frac{1}{\gamma_{t0}}\ln\frac{\Lambda^2}{\sqrt{\gamma_{t0}^2+\mu^2}}-\frac{\mu}{\gamma_{t0}^2+\mu^2+ U \gamma_{t0}}\Bigr\}(1-R_0^\prime),
\end{equation*}
where we have neglected the term $\propto \Lambda^2\gamma_{t0}^\prime/\gamma_{t0}^2$ and the term $\propto \frac{\mu}{\gamma_{t0}^2}\gamma_{t0}^\prime\ln\frac{\Lambda^2}{\sqrt{\gamma_{t0}^2+\mu^2}}$, both of which give contributions smaller than $\alpha_{xx}^R$. For $|\mu|\sim \gamma_{t0}$, $\alpha_{xx}^L\sim \frac{1}{ U }\ll \alpha_{xx}^R$ where $\alpha_{xx}^R\sim \frac{\mathcal{F}_{R0}}{ U ^2}\frac{1}{(\ln\frac{\Lambda^2}{ U })^3}$. For $|\mu|\gg \gamma_{t0}$, $\alpha_{xx}^L\sim \frac{1}{ U }\ll \alpha_{xx}^R$ where $\alpha_{xx}^R\sim \frac{\mathcal{F}_{R0}}{ U |\mu|}\frac{1}{(\ln\frac{\Lambda^2}{|\mu|})^2}$. However, since $\gamma_{t0}^\prime=0$ for $\mu=0$, $\alpha_{xx}^L$ is potentially important for $|\mu|\ll\gamma_{t0}$. In this limit, $\alpha_{xx}^L\sim \frac{1}{ U }$. We note that in the normal metallic state, $\alpha_{xx}^N\sim \frac{\mathcal{F}}{\mathcal{D}}\frac{1}{ U }$, and since $\mathcal{F}/\mathcal{D}\sim \mathcal{O}(1)$ for a generic dispersion $\varepsilon_\mathbf{p}$, $\alpha_{xx}^L\sim \alpha_{xx}^N$. We thus argue that $\alpha_{xx}^L$ can be neglected for the calculation of the singular contribution to $\alpha_{xx}$.

\end{document}